\documentclass[12pt]{article}
\usepackage{amssymb,latexsym,graphics,epsfig}

\begin{document}

\begin{center}
{\bf \Large Influence of Expansion on Hierarchical Structure}\\
\vspace*{0,5cm}
{\bf \large Bruce N. Miller
\footnote{Dept. of Physics and Astronomy, Texas Christian University,
Fort Worth, Texas 76129, U.S.A.\\
B.Miller@tcu.edu}
{\bf \large J.L. Rouet
\footnote{Laboratoire de Math\'ematique, Applications et Physique 
Math\'ematique - UMR 6628, Universit\'e d'Orl\'eans, UFR des Sciences, 
F-45067 Orl\'eans Cedex 2, France\\
jean-louis.rouet@univ-orleans.fr}}}\\
\vspace*{0,25cm}
\end{center}

\paragraph{Abstract}
We study a one dimensional model of gravitational instability in an
Einstein-de Sitter universe. Scaling in both space and time results in
an autonomous set of coupled Poisson-Vlasov equations for the field and
phase space density, and the $N$-body problem. Using dynamical
simulation, we find direct evidence of hierarchical clustering. A
multi-fractal analysis reveals a bifractal geometry similar to that
observed in the distribution of galaxies. To demonstrate the role
of scaling, we compare the system to other one dimensional
models recently employed to study structure formation. Finally we show
that the model yields an estimate of the time of galaxy formation of
the correct order.\\


The discovery nearly twenty five years ago of large scale structures in
the universe \cite{Peebles} stimulated several new approaches to
cosmology. In order to explain, or at least model, the hierarchical
distribution of galaxies in clusters, and clusters of clusters,
surrounding immense pancake shaped voids, it was proposed that matter
in the universe occupys a fractal set, with a definite fractal geometry
and dimension \cite{Peebles,Man}. It is currently believed that the
process for creating this singular distribution of mass follows from
the hydrodynamic flow of dark matter, either hot or cold
\cite{Peebles,Zel}. In the former case structure evolves in the top
down mode, while the latter is bottom up. Both of these themes
introduce their own peculiar set of difficulties~: the fractal
assumption plays havoc with the homogeneity at large scales required by
the standard model \cite{Peebles} while the bottom up scenario may be
problematic because it predicts galaxy formation at a time when the
background density is small \cite{Peebles2}.

Before (and after) the discovery of large scales, useful information
was obtained by assembling statistical information concerning the
distribution of galaxies in the sky. The primary tool for organizing
this information is the construction of correlation functions
representing the distribution of pairs of galaxies, triples, etc\dots
While complete functions at all orders are unobtainable, specific
symmetries and scaling laws have emerged from the data. First the
validity of a power law was established for the dependence on
inter-galactic separation of the pair correlation function, with
exponent $\gamma=1.8$ \cite{Peebles}. Second, it was found that the
data on higher order distributions supported the assumption that the
Nth order correlations are homogenious functions of order
$(N-1)\gamma$ \cite{Peebles}.

In pioneering work, Bailin and Schaeffer \cite{Bai} employed these
properties to reconcile the assumption of homogeneity on large scales
with fractality on intermediate and small scales. They demonstrated
that the distribution of galaxies was not consistent with a normal
fractal, but rather with a geometry which is characterized by two
length scales, $l_1$ and $l_2$, where $l_o>>l_1>>l_v$, and
$l_v>>l_2>>l_c$. Here the transition to homogeneity occurs at $l_o$,
the maximum size of voids is $l_v$, and the typical size of clusters
is $l_c$. They named this object ``bifractal''. More recently the
approach has been successfully applied to the distribution of mass in
halos \cite{Val}.

In addition to the Hausdorff dimension, the existence of a hierarchy of
generalized dimensions, $D_q$, which are closely related to the
properties of correlation functions at small scales for $q\geq{2}$, is
now well established \cite{Tel}. In particular, $D_0$ is simply the box
counting dimension, $D_1$ is the information dimension, and $D_2$ is
the correlation dimension. For a normal fractal with self-similar
geometry at all scales, $D_{q+1}\leq{D_q}$ \cite{Tel}. An unusual
feature of bifractal geometry is the violation of this inequality,
which can be taken as its signature \cite{Bai,Val}.

A central question for astrophysics is whether it is possible to
construct a single, consistent, dynamical model which (1) obeys
physical law, (2) is homogenious on large scales, (3) exhibits
hierarchical clustering or agregation, and (4) is characterized by
bifractal geometry.  As early as the 1930's, dynamical models of cosmic
evolution were introduced which joined the hubble flow with the
gravitational field \cite{Peebles}. They were later rejected as
possible sources of structure formation because, in the linear
approximation, density fluctuations about a homogenious background did
not grow sufficiently rapidly to produce galaxies during the lifetime
of the universe \cite{Peebles}. Here we revisit a version of one of
these models with the goal of following the evolution of small initial
fluctuations over the full nonlinear regime.

In the following we introduce a version of a de Sitter universe obeying
classical dynamics appropriate to the post-recombination epoch. As in
the relativistic Tolman-Bondi models \cite{Silk}, we simplify the
geometry by assuming spherical symmetry about an observer. In addition,
we examine a region sufficiently distant from the observer that the
effects of curvature can be ignored. In contrast with the earlier
treatments \cite{Peebles,Silk}, we rescale both space and time to
obtain a completely autonomous dynamical system. The model was first
introduced by Rouet and Feix \cite{Rouet1,Rouet2}, who showed that
agregation was stimulated by excitations at the Jeans length, and
computed a box counting dimension less than unity, suggesting fractal
behavior. Here we use Vlasov-Poisson theory to characterize the central
properties of the dynamics. We then use dynamical simulation to examine
the consequences of a variety of initial conditions. We carry out a
complete dimensional analysis of the phase plane and density
distribution as time evolves, and show that the system exhibits
bifractal geometry in all cases where the Jeans length is initially
available to the fluctuations. We demonstrate that the earliest time
for agregation is nearly independent of initial conditions or
population. We use the value of this scaled time to estimate the
earliest epoch for the appearance of galaxies in the universe, with
surprising results.\\


Consider a spherically symmetric, homogenous, expanding universe with
density $\rho(t)$ under conditions where Newtonian mechanics applies.
Let $C(t/t_0)$ be the scale factor, so that the distance $l(t)$ between
two objects at the time $t$ is related to that at the earlier time
$t_0$ by $l(t)=C(t/t_0)l(t_0)$. Here $t_0$ does not signify the big
bang, but rather an arbitrary, later, time where only gravitational
phenomena play an important role in the cosmic evolution.  To this
expansion we must add a residual motion which is a small perturbation
of the Hubble flow, but leads the system to a nonlinear regime.

From spherical symmetry, we only need to track a single coordinate, the
radius. Thus our system elements are represented by concentric mass
shells. The description can be further simplified by assuming that we
are far from the center of symmetry and that the length of the system
is small compared to the radius of the spheres, so that we may replace
the shells with planar sheets. Then the equation of motion of a sheet
with coordinate x is simply

\begin{equation} m\frac{d^2x}{dt^2}=E(x,t) \label{eq1} \end{equation}

\noindent where $E$ is the gravitational field.

For the special case of an Einstein-de Sitter universe, there is a
unique rescaling of space and time to a new frame in which the
dynamical evolution is autonomous. Introduce new coordinates $\hat{x}$
and $\hat{t}$

\begin{equation}
x=C(t)\hat{x} ,\quad  dt=A(t)^2d\hat{t} 
\label{eq2}
\end{equation}

\noindent and scale the field to keep the form of Poisson's equation
invariant. To insure that the transformed version of eq(\ref{eq1}) is
autonomous, we must choose $C=(t/t_0)^{2/3}$ and $A(t)=(t/t_0)^{1/2}$.
The complete three dimensional expansion is taken into account with
this choice of $C(t)$. In the transformed frame, the density
$\hat{\rho}$ is constant. Following standard practice, we choose the
inverse Jeans frequency, defined by $\hat{\omega}_j^2=4\pi G
\hat{\rho}$, where $\hat{\rho}$ is the density, as our unit of time
(See \cite{Rouet1} for a more complete discussion). The equation of
motion in the new frame then takes the form

\begin{equation}
\frac{d^2\hat{x}}{d\hat{t}^2}+\gamma\frac{d\hat{x}}{d\hat{t}}-\hat{x}=\hat{E}.
\label{eq5} \end{equation}

\noindent where the choice of a neutralizing background requires
$\gamma=1/\sqrt{2}$. Equation (\ref{eq5}) describes the motion of a
collisionless system of particles moving under their mutual
gravitational field. From Gauss' law applied to uniform mass sheets,
the field experienced by a particle on the line is simply proportional
to the net difference in mass of the particles to its right and left.
The transformations have induced both a linear friction and a constant,
``negative mass'', background density $\rho_b$. Thus the system is
equivalent to a single component plasma with a drag force in which
opposite charges repel and like charges attract. In the following we
will drop the decorations on x and t, with the understanding that we
are in the rescaled frame.

In the Vlasov limit, the system is amenable to a continuum description.
Usefull information can be had from the time dependant Vlasov equation,
fixing the evolution of the density in the $x-v$ phase plane. For
example, we easily find that the system energy decreases at a rate
proportional to the kinetic energy, while the entropy decreases at the
constant rate $-2\gamma$, and the Tsallis entropy decreases
exponentially for $q>1$. This tells us that the mass is being
concentrated in regions of decreasing area of the phase plane,
suggesting the development of structure. By asserting a Euclidean
metric in the phase plane, we can also investigate local properties
such as the directions of maximum stretching and compression, as well
as the local vorticity. We find that the rate of separation between two
nearby points is a maximum in the direction given by

\begin{equation} \tan(2\theta)=(1+\rho + \rho_b)/\gamma, \label{eq9}
\end{equation}

\noindent where $\theta$ defined a slope in phase space. Thus, in
regions of low density, we expect to see lines of mass being stretched
with constant positive slope in the phase plane.\\


For a descrete population, the dynamics can be viewed as a sequence of
particle crossings. Between an adjacent pair of crossings, Eq
(\ref{eq5}) can be integrated analytically to yield an explicit
solution for the position and velocity of each particle. Following the
selection of an initial condition, an event driven algorithm was
employed to compute the crossing sequence. The details of the algorithm
are described elsewhere \cite{Rouet1}. The evolution was followed until
boundary effects became noticeable, typically in 15-20 dimensionless
time units.  In a few instances (see below) much longer simulations
were carried out. The evolution of the system was investigated with
dynamical simulation for system populations of 10,000 and 50,000
particles (or sheets).

Depending on which cosmology we select, the linear regime is
characterized as Gaussian, $1/f$ noise, or a Brownian motion of the
normal or fractional variety \cite{Peebles}. In order to judge the
robustness of the dynamics, the response of the system to several
initial conditions was investigated. Here we discuss the two extremes:
an initial Gausssian distribution in velocity, and a Brownian motion in
position. In all cases, the particle positions were initally located
equidistantly along the coordinate line. For the Gaussian (or
isothermal) case, the velocity of each particle was independently
selected from a normal distribution of mean zero and variance
$\sigma_0^2$. The initial temperature was chosen such that the length
of the system was about 2,000 times the  Jean's length,
$\lambda_j=\sigma_0\omega_j$. In contrast, for the Brownian motion, the
increment in velocity from one particle to another along the line is
normally distributed. Thus, in contrast to the isothermal distribution,
the initial velocities of neighboring particles are strongly
correlated.

\begin{figure}[!ht] 
\centerline{\includegraphics[width=12cm]{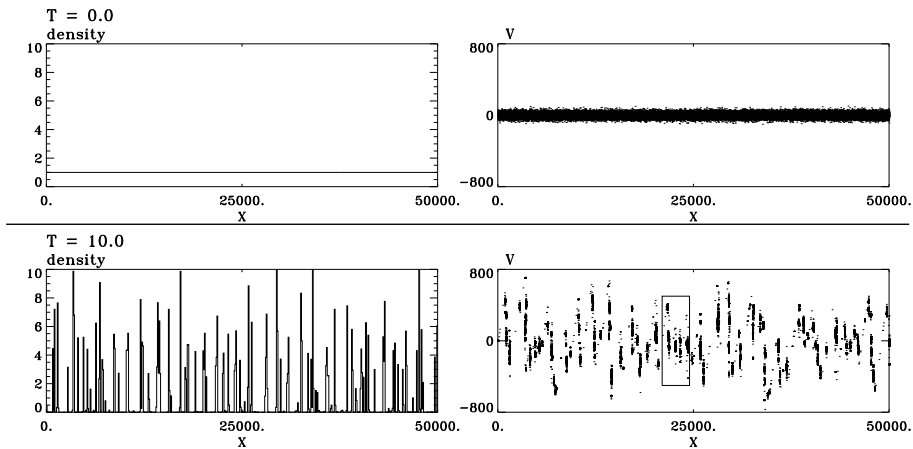}}
\caption{\label{fig1} Density and phase plane distribution for the
initial condition, and after evolution for ten dimensionless time
units, ($\hat{\omega}_j\hat{t}=10$ for $N=50,000$ particles. The
velocities are initially chosen at random from a Gaussian distribution
with initial variance $L/\lambda_j=2,000$.}
\end{figure}

As time evolved, inspection of the distribution of the cloud of points
in the phase plane, and their positions on the line, were similar for
each initial condition, and we display the Gaussian (see Fig
\ref{fig1}). For short times, before crossings can occur, the field
experienced by each particle is very weak, and we observe the
exponential decrease in speed induced by the friction. However, as the
number of crossings increases, the effects of instability become
apparent. Typically, in about four dimensionless time units, two types
of structure become obvious - lines and clumps. In the low density
regions, particles are distributed along a line of constant slope in
the phase plane, as suggested by the stretching analysis discussed
above, while in the high density regions they form clumps of roughly
equal size. As the simulation goes forward in time, the process is
repeated in hierarchical fashion, i.e. the clumps merge into bigger
clumps. To test the role of Jean's length, we also prepared a much
hotter system, where the Jean's length exceeded the system size. In
that case, after the system cooled, a single clump formed near one
boundary. However, even this distribution evidenced hierarchical
layering around its center in the phase plane.\\

\begin{figure}[!ht]
\centerline{\includegraphics[width=8cm]{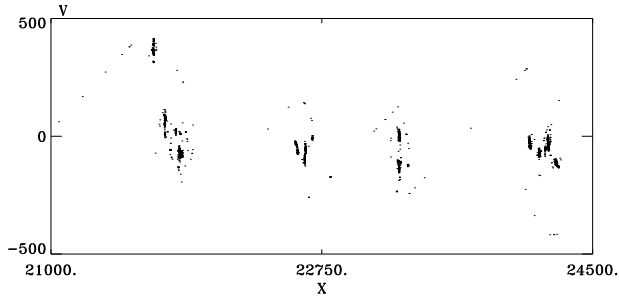}}
\caption{\label{fig1b} The zoom of an area in the phase plain (the
insert in figure \ref{fig1}) containing one of the agregates formed at
$\hat{\omega}_j\hat{t}=10$ emphasizes the hierarchical structure.}
\end{figure}


It is natural to assume that the apparently self similar structure
which develops in the phase plane as time evolves has fractal geometry,
but we will see that things aren't so simple. An earlier study found a
box counting dimension of about 0.6 for an initial waterbag
distribution (uniform on a rectangle in the phase plane) \cite{Rouet2}.
Since the structures which evolve are strongly inhomogenious, to gain
further insight we decided to carry out a multi-fractal analysis
\cite{Tel} in both the phase plane and the position coordinate. To
accomplish this we partitioned each space into cells of length $l$. At
each time of observation in the simulation, a measure $\mu_i=N_i(t)/N$
was assigned to cell $i$, where $N_i(t)$ is the population of cell $i$
at time $t$ and $N$ is the total number of particles in the simulation.
The generalized dimension of order $q$ is defined by \cite{Tel}

\begin{equation}
D_q=\frac{1}{q-1}\lim_{l \rightarrow 0}\frac{\ln{C_q}}{\ln {l}},
\quad C_q=\Sigma \mu_i.
\label{eq10} 
\end{equation}

\noindent where $D_0$ is the box counting dimension, $D_1$, obtained
by taking the limit $q\rightarrow1$, is the information dimension, and
$D_2$ is the correlation dimension \cite{Tel}. As $q$ increases above
$0$, the $D_q$ provide information on the geometry of cells with
higher population. If it exists, the scaling range of $l$ is defined
as the interval on which plots of $\ln C_q$ versus $\ln l$ are
linear. Of course, for the special case of $q=1$, we plot
$\Sigma\mu_i\ln\mu_i$ vs. $\ln l$. If a scaling range can be found,
$D_q$ is obtained by taking the apropriate derivative. It is well
established by proof and example that, for a normal, homogenious,
fractal, all of the generalized dimensions are equal, while for an
inhomogenious fractal, e.g. the Henon attractor, $D_{q+1}\leq{D_q}$
\cite{Tel}.

\begin{figure}[!ht] 
\centerline{\includegraphics[width=8cm]{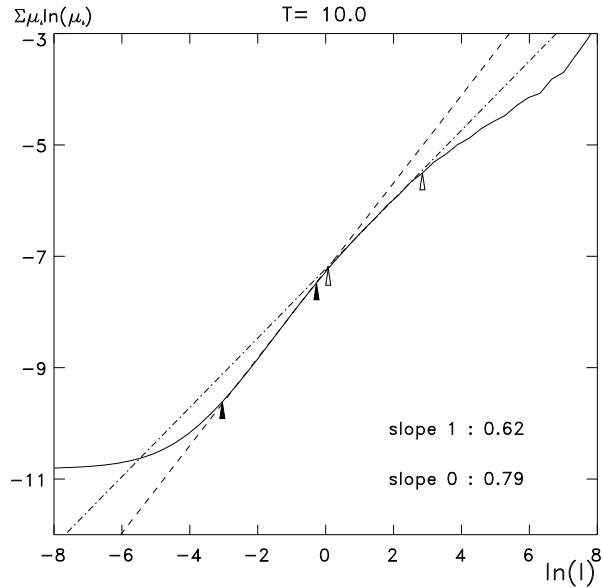}}
\caption{\label{fig2} Plot of $\Sigma \mu_i \ln(\mu_i)$ versus the log
of the box size. The two dashed lines show the two regions for which
the curve reveals a linear scaling regime. For smaller and larger
scales (not shown) the slope takes on the obvious values of 0, and 1,
respectively.}
\end{figure}

\begin{figure}[!ht]
\centerline{\includegraphics[width=8cm]{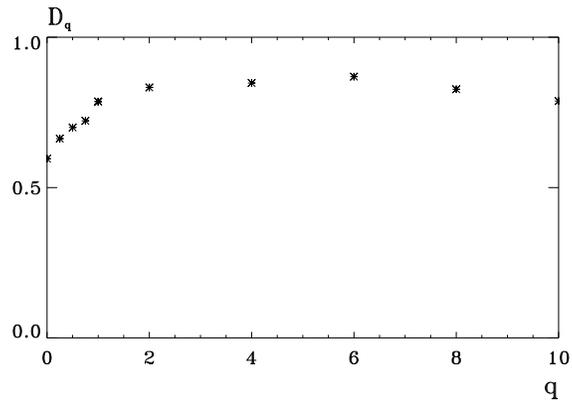}}
\caption{\label{fig3} Plot of $D_q$ versus $q$ for the same initial
conditions and time as Fig.\ref{fig2}.}
\end{figure}

As expected, initially, and for a short time afterwards, all
simulations showed a box counting dimension of two in the phase plane,
and one along the coordinate axis. As time progressed, however, for
each of the two initial conditions discussed above, at least one clear
scaling range developed early in the simulation. For both the Gaussian
and the Brownian Motion, $D_0$ quickly converged on about 0.6 and
remained there for most of the simulation.  The size of the scaling
range depended on both the ellapsed time into the simulation and the
value of $q$. We started our investigation by computing the first three
dimensions. We were surprised to observe that, in fact, $D_2>D_0$ in
all cases~! Moreover, for $q\geq1$, a secondary, weaker, scaling range
was also detected. In Fig. \ref{fig2} we plot $C_1$ versus $\ln l$ at
the time $T=10$ for the isothermal initial condition. We clearly see
one dominant scaling range for small $l$, a second scaling range for
intermediate $l$, and the possibility of a third range for larger $l$.
It is suggestive that the transition between the first two scaling
regimes occurs roughly at the Jeans length of the initial distribution.
Since the size of the first clusters are approximately equal to the
Jeans length, the suggestion is that the fractal geometry within the
clusters differs from that of the less populous ``voids''. In
Fig.\ref{fig3} we plot $D_q$ versus q for the same conditions and time
as Fig. \ref{fig2}. We see that Most of the change in dimension occurrs
when $0<q<1$ . Although there is little change in $D_q$ for $2<q$, the
dominant scaling range grows progressively smaller with increasing $q$.
This type of behavior was first inferred in a study of the observed
correlations of galaxy positions by Bailin and Schaeffer \cite{Bai} who
named the geometry bifractal.\\

In the last few years, the dynamics of a group of autonomous one
dimensional models has been studied for the purpose of gaining new
insight concerning the development of hierarchical structures. These
include the adhesion model \cite{Adhe}, Burgers Equation \cite{Bur},
and different versions of the system studied here, either with no
scaling \cite{Koy} (neither background nor friction), incomplete
scaling \cite{Aur}(background but no friction), or with fractal initial
conditions \cite{Tay}. In the adhesion model, particles move on the
line according to their mutual gravitational attraction. However, when
they cross, they stick. In this system agregation into a single large
cluster occurs quickly, but a finite fraction of the system remains
associated with smaller clumps for a long time\cite{Adhe}. Burgers
equation has been carefully studied for a range of initial conditions
which vary according to the correlation of initial velocities on the
line \cite{Bur}. For some initial states, shocks develop, yielding
velocity portraits $v(x)$ similar to what we see in Fig. 1\ref{fig1}.

Very recently, structure formation was observed in the conservative one
dimensional gravitational system, both with and without background
\cite{Koy,Aur,Tay}. To get a better sense of how scaling influences the
development of structure, we also performed simulations of these
systems and examined their multifractal properties. The results were
interesting: Similar, hierarchical, structures developed in each
system. However, the generalized dimensions were larger in each case,
about 0.8 for $D_0$ , and bifractal behavior was much weaker than in
the dissipative version studied here. In fact, with no background, we
did not observe a bifractal structure and, with the background present,
although we found $D_2>D_0$, the difference was small. In each of these
systems, the dimensions were less stable and varied with time. In the
case without background, the fractal appearance washed out with the
subsequent virialization. With the background present, the structure
endured for a longer time. For contrast, we carried out a long
simulation of the fully scaled system. We found that fractal structure
and scaling endured long after the system retreated to a single cluster
confined to a small region of the phase plane near the right hand
boundary.\\


We observed earlier that, so long as the Jeans length was initially
accessible to the system, the formation of structure occurred rapidly
and robustly at about 4 dimensionless time units for all attempted
initial conditions. Of course, this was in scaled time. Converting back
to cosmic time, we simply find $t=t_o\exp{(3/2)(\hat{t}-\hat{t_0})}$. 
If we take $t_o$ as the time of recombination in a de Sitter universe,
approximately $10^5-10^6$ years \cite{Peebles}, and scaled time
$\hat{t}-\hat{t_0}=4.0$, we obtain a time in the range 0.5- 5 billion
years for the appearance of the first galaxies. It may seem pretentious
to use such a simple model to identify the time when galaxies first
appear.  However, we should not be surprised that what we find is of
the correct order since the model consistently couples gravity with
expansion, and dynamical simulation shows that the results are similar
under a large variety of inital conditions. More complete theories will
manifest more accurate details, but the overiding source of structure
is nonlinearity, and this wont change.   

An interesting, and potentially useful, feature of the model is that it
unambiguously exhibits what has been coined bifractal geometry. While
this type of structure has been inferred from correlation functions for
the distribution of galaxies, being able to construct the geometry with
an autonomous gravitational system could yield real benefits in the
future.  In a larger work we will elucidate the multifractal features
in more detail and study their connection with correlations in position
and in the phase plane. It is anticipated that the success of this
approach will stimulate the study of the nonlinear evolution of more
realistic cosmological models in higher dimension.

\end{document}